\documentclass[twocolumn]{aastex6}
\usepackage{graphicx}

\usepackage[caption=false]{subfig}

\begin{document}

\title{How Do We Optimally Sample Model Grids of Exoplanet Spectra?}

\author{Chloe Fisher\altaffilmark{1,2}}
\author{Kevin Heng\altaffilmark{1,3,4}}

\altaffiltext{1}{University of Bern, Center for Space and Habitability, Sidlerstrasse 5, CH-3012, Bern, Switzerland; chloe.fisher@unibe.ch}
\altaffiltext{2}{University of Oxford, Department of Physics, Denys Wilkinson Building, Oxford OX1 3RH, UK}
\altaffiltext{3}{University of Warwick, Department of Physics, Astronomy and Astrophysics Group, Coventry CV4 7AL, UK}
\altaffiltext{4}{Ludwig Maximilian University, University Observatory Munich, Scheinerstrasse 1, Munich D-81679, Germany}

\begin{abstract}
The construction and implementation of atmospheric model grids is a popular tool in exoplanet characterisation. These typically vary a number of parameters linearly, containing one model for every combination of parameter values. Here we investigate alternative methods of sampling parameters, including random sampling and Latin hypercube (LH) sampling, and how these compare to linearly sampled grids. We use a random forest to analyse the performance of these grids for two different models, as well as investigate the information content of the particular model grid from \cite{goyal19}. We also use nested-sampling to implement mock atmospheric retrievals on simulated JWST transmission spectra by interpolating on linearly sampled model grids. Our results show that random or LH sampling out-performs linear sampling in parameter predictability for our higher dimensional models, requiring fewer models in the grid, and thus allowing for more computationally intensive forward models to be used. We also find that using a traditional retrieval with interpolation on a linear grid can produce biased posterior distributions, especially for parameters with non-linear effects on the spectrum. In particular, we advise caution when performing linear interpolation on the C/O ratio, cloud properties, and metallicity. Finally, we find that the information content analysis of the grid from \cite{goyal19} is able to highlight key areas of the spectra where the presence or absence of certain molecules can be detected, providing good indicators for parameters such as temperature and C/O ratio. 
\end{abstract}


\section{Introduction}

With the recent launch of the James Webb Space Telescope (JWST) and the upcoming developments in ground-based observatories, we are stepping into a new era of exoplanet data. These new facilities promise an explosion in the precision and sensitivity of spectra of exoplanet atmospheres, which will require a matching advancement in our analysis techniques. The current state-of-the-art in exoplanet analysis uses atmospheric retrieval to search parameter space for the model that best fits the data \citep[e.g.][]{madhusudhan09,benneke13}. These traditionally employ a Bayesian sampling algorithm such as an MCMC or nested-sampling, in conjunction with an atmospheric model. In a single retrieval, tens of thousands of models are computed on-the-fly and compared to the data, meaning we are inherently limited in the complexity of the physical models we can use. Most retrievals rely on 1-dimensional atmospheric models, with some recent work branching out into forms of 2-D models \citep[e.g.][]{irwin20}. However, many studies have investigated potential biases in the results from 1-D retrievals, and the detrimental effects these can have when attempting to accurately characterise an exoplanet \citep[e.g.][]{feng16,line16,taylor20}. With improved data from upcoming instruments such as JWST and the ELT, these biases will only worsen.

An alternative method of exoplanet analysis involves the computation of grids of atmospheric models, constructed by varying each parameter in turn. These grids have fewer computational restrictions, allowing for more complex physics to be included in the model. The linear structure of the grids enables one to study the individual effects of each parameter on the spectrum and assess the sensitivity of observations. The grids can also be used to exclude particular models when analysing data. For example, \cite{dewit18} are able to rule out hydrogen-dominated atmospheres for the Trappist-1 planets simply by visual inspection. In more recent years, several groups have developed techniques that use machine learning to perform atmospheric retrieval and other analyses by training on a set of synthetic spectra \citep[e.g.][]{waldmann16,marquez-neila18,zingales18,cobb19,fisher20,matchev22,ardevolmartinez22}. This form of retrieval allows one to use model grids provided by other groups, without requiring access to the original model code or relying on an interpolation method. Most model grids are either open-source or can be provided on demand, and range from brown dwarf spectra \citep{burrows97,allard01,allard14,marley21} to Global Circulation Models \citep{edson11,perna12,tan19,beltz21} to exoplanet spectra \citep{fortney10,molliere17,kempton17,goyal18,goyal19,goyal20}. This presents an interesting opportunity for machine learning retrievals to take advantage of these grids, and provide some comparison across different models. This was investigated in a study of brown dwarfs \citep{oreshenko20}, which compared grids from three different groups to highlight differences in the models. 

The linear spacing of these model grids has one key disadvantage -- as the number of parameters increases, the number of models required increases exponentially and becomes prohibitive. An advantage of using machine learning is that it is able to automatically disentangle the parameter effects by learning from a large number of examples. This suggests that the linearly sampled grid is sub-optimal for machine learning retrievals. The use of stratified sampling methods, a type of random sampling that ensures each subdivision (or \textit{strata}) of parameter space is evenly sampled, have been demonstrated to optimise computer experiments for many years \citep[e.g.][]{mckay79,wang03,chalom12}. Therefore, one would expect random and stratified sampling methods to outperform linear sampling in the problem of machine learning atmospheric retrievals. In this paper we investigated to what degree this applies. We test different methods of sampling exoplanet model grids for various types of analysis. We create our own model grids with an increasing number of parameters and different sampling methods, and then compare the predictability of each parameter using the random forest \citep{marquez-neila18}. We also consider the model grid from \cite{goyal19}, and use an analytical approximation to create differently sampled versions. We then use the random forest to analyse the grids for different purposes.

\section{Methods}

\subsection{Modelling}

Here we describe the methods and assumptions used to generate our grids of atmospheric models. 

\subsubsection{Analytical Model}
\label{sec:analytical_model}

In order to test sampling methods and grid sizes, we use a simplified analytical model, assuming an isothermal, isobaric atmosphere. This follows the work of \cite{lecavelierdesetangs08,dewit13,betremieux17,heng17a,jordan18,heng19,fisher18}, and allows one to write down an analytical expression for the transit radius, given by equation (2) in \cite{fisher18}. The atmospheric opacity is given by 
\begin{equation}
    \kappa = \sum_{i} \frac{X_i m_i \kappa_i}{m} + \kappa_{\rm CIA} + \kappa_{\rm haze} + \kappa_{\rm cloud},
\end{equation}
where $m$ is the mean molecular mass, and $X_i$, $m_i$, and $\kappa_i$ are the volume mixing ratio, mass, and opacity of species $i$, respectively. $\kappa_{\rm CIA}$ is the opacity associated with collision-induced absorption (both H$_2$-H$_2$ and H$_2$-He), taken from \texttt{HITRAN} \citep{rothman13}. $\kappa_{\rm haze}$ and $\kappa_{\rm cloud}$ follow different equations in two different models we consider (see Sections \ref{sec:freechem_models} and \ref{sec:goyal_models}), but are generally associated with Rayleigh scattering and a gray cloud, respectively. The cross-section due to Rayleigh scattering is taken from \cite{vardya62}.

To mimic spectra we expect to obtain from JWST's NIRSpec Prism mode, we bin our models to $\sim$400 points in the range $0.6-5.3\mu$m, giving a resolution of $\sim$100. We then add random Gaussian noise, assuming the uncertainty on each spectral point to be 20ppm.

\subsubsection{Free Chemistry Models}
\label{sec:freechem_models}

For our first set of models we assume free chemistry. This means the abundances of each molecule can take any value, which allows for a greater freedom in the models, but could lead to unphysical compositions. For these models, we include a varying subset of the molecules H$_2$O, CO, CO$_2$, CH$_4$, C$_2$H$_2$, HCN, and NH$_3$. The opacities for these molecules are computed using the open-source \texttt{HELIOS-K} opacity calculator \citep{grimm15,grimm21}, and the line lists are taken from the \texttt{ExoMol}, \texttt{HITRAN} and \texttt{HITEMP} databases -- \cite{polyansky18} (H$_2$O); \cite{li15} (CO); \cite{rothman10} (CO$_2$); \cite{yurchenko14} (CH$_4$); \cite{gordon17} (C$_2$H$_2$); \cite{barber14} (HCN); \cite{yurchenko11} (NH$_3$). The molecular opacities are sampled every 0.01 cm$^{-1}$ in wavenumber space, at a pressure of 1 mbar. In each set of models, we vary the temperature and molecular abundances of the included species. For one version of this set we also include a non-gray cloud model, following equation 9 of \cite{fisher18},
\begin{equation}
    \kappa_{\rm cloud} = \frac{\kappa_0}{Q_0 x^{-a} + x^{0.2}}.
\end{equation}
This analytical cloud model comes from \cite{kitzmann18}. In this work, we vary three of the parameters -- the factor $\kappa_0$,  the index $a$, and the cloud particle size $r_c$ (measured in cm). The cloud composition $Q_0$ is set to 50, since previous studies have shown it to be unconstrained in retrievals \citep[e.g.][]{fisher18}. For this model, $\kappa_{\rm haze}$ is simply opacity due to Rayleigh scattering. The range of values spanned by the grids for all the possible parameters in our free chemistry model are shown in Table \ref{tab:priors}.

For all the models we assume the same planetary parameters as WASP-12 b: $R_{\rm p}=1.79R_{\rm J}$, $g=977$cm s$^{-2}$, $R_{*}=1.57R_{\odot}$. 

\begin{table}
\begin{center}
\caption{Prior ranges for all possible parameters in our free chemistry model.}
\label{tab:priors}
\vspace{0.1in}
\begin{tabular}{|l|c|c|}
\hline
    Parameter & Prior Range \\
    \hline
    T (K) & [500, 2900] \\
    $\log{X_{i}}$ & [-13, -1] \\
    $\log{\kappa_0}$ & [-10, -1] \\
    $a$ & [3, 6] \\
    $\log{r_c}$ & [-7, -1] \\
    \hline
\end{tabular}
\end{center}
\end{table}

\subsubsection{Goyal Models}
\label{sec:goyal_models}

Our second set of models emulates the grid from \cite{goyal19}, and thus are termed ``Goyal models". \cite{goyal19} present a scalable grid of exoplanet transmission spectra, varying the temperature, gravity, metallicity, C/O ratio, haze, and cloud. Their models are computed using \texttt{ATMO} -- a 1-D radiative-convective-equilibrium model for planetary atmospheres \citep{amundsen14,tremblin15,tremblin16,drummond16}. They use isothermal $P-T$ profiles, assuming chemical equilibrium. Full details on their implementation of the model can be found in the paper.

We simulate these models using the analytical model described in Section \ref{sec:analytical_model}. To implement equilibrium chemistry in our model, for a given temperature, metallicity, and C/O ratio, we use the validated analytical model of \cite{heng16b} that includes carbon, oxygen and nitrogen. From this we obtain abundances for H$_2$O, CO, CO$_2$, CH$_4$, C$_2$H$_2$, HCN, and NH$_3$. The opacities used for these molecules are the same as in Section \ref{sec:freechem_models}.

In the grid from \cite{goyal19}, the parameters are sampled as follows. The temperature is sampled every 100 K, from 300 to 2600 K. The surface gravity can take one of the four values 5, 10, 20, or 50 ms$^{-2}$. The atmospheric metallicity controls the elemental abundances, including oxygen, and is sampled at 0.1, 1, 10, 50, 100, and 200 times solar. The C/O ratio controls the carbon abundance, and takes the values 0.35, 0.56, 0.7, and 1.0. The haze parameter controls small scattering aerosol particles, and is implemented as $\alpha_{\rm haze}$ in the equation $\sigma(\lambda)=\alpha_{\rm haze}\sigma_0(\lambda)$, where $\sigma(\lambda)$ is the total scattering cross-section, and $\sigma_0(\lambda)$ is the H$_2$ Rayleigh scattering cross-section. $\alpha_{\rm haze}$ is sampled at 1, 10, 100, and 1100. The cloud is treated as large particles with a gray opacity, and the parameter is implemented as $\alpha_{\rm cloud}$ in the equation $\kappa(\lambda)_c=\kappa(\lambda) + 2\kappa_{\rm H_2}\alpha_{\rm cloud}$, where $\kappa(\lambda)_c$ is the total scattering opacity, and $\kappa_{\rm H_2}$ is the scattering opacity due to H$_2$ at 350 nm, which \cite{goyal19} states as $\sim2.5\times10^{-3}$ cm$^2$g$^{-1}$. $\alpha_{\rm cloud}$ is sampled at 0, 0.06, 0.2, and 1.0. This leads to a total of 36,864 models.\footnote{Note that some of the parameter values in the open-source grid have changed since \cite{goyal19} was published, and the information on this can be found in the \texttt{readme.txt} file in their google drive.}.

\subsubsection{Analytical vs Full Model Comparison}
\label{sec:model_comparison}

\begin{figure*}
    \centering
    \includegraphics[width=\textwidth]{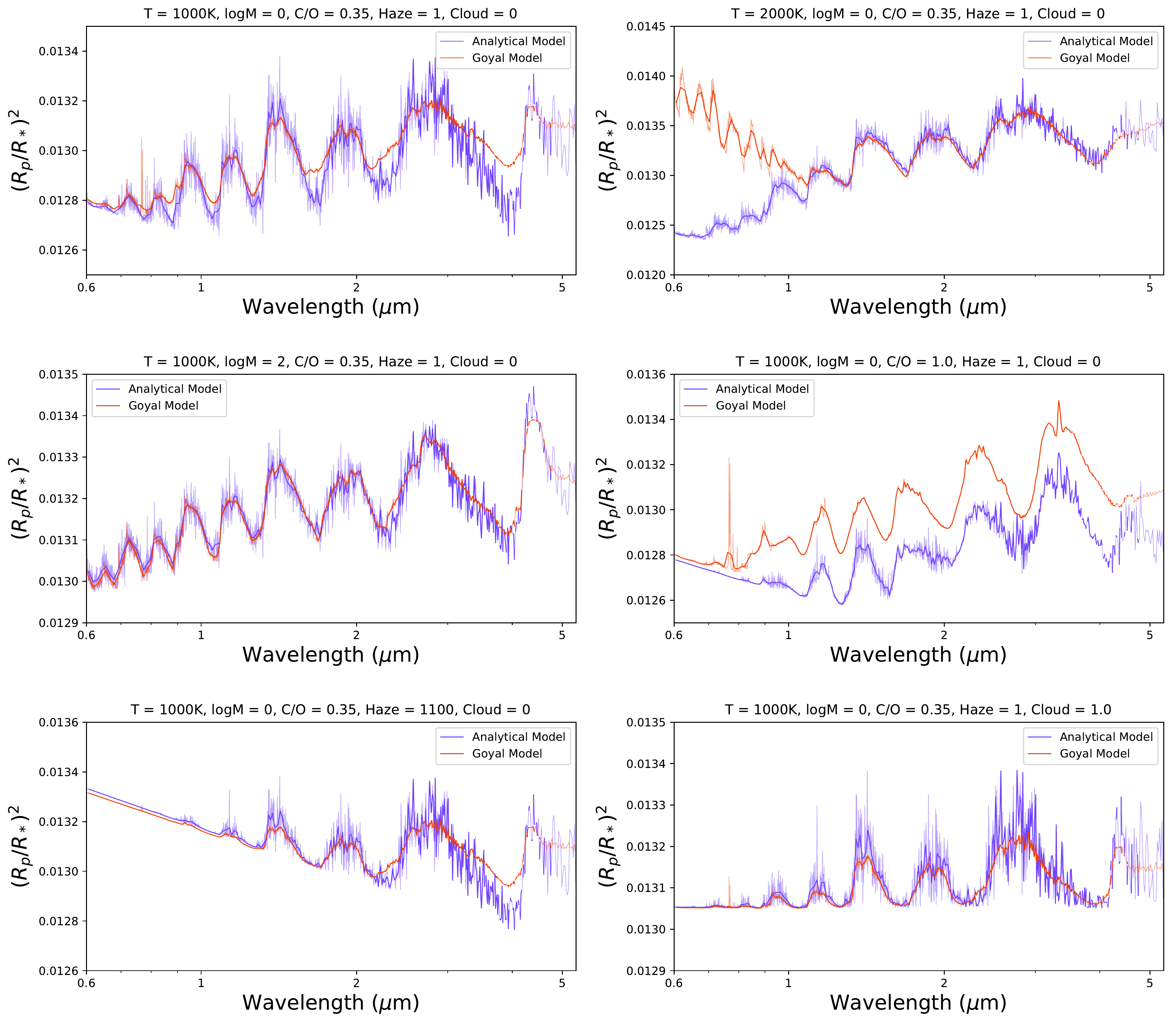}
    \caption{Comparison of our analytical models with the corresponding models from \cite{goyal19}, shown for different parameter values. The analytical model is shown in blue and the model from \cite{goyal19} is shown in red.}
    \label{fig:spectra_comparison}
\end{figure*}

There are several key approximations in our analytical model, such as the isobaric atmosphere and constant chemical abundances. Figure \ref{fig:spectra_comparison} shows a comparison of six models with their corresponding model from \cite{goyal19}. Each model has only one parameter changed with respect to the top left model. The top right model shows the effect of a higher temperature. In this case, the absorption due to TiO and VO dominates in the bluer wavelengths for the \cite{goyal19} model. Since we don't include TiO or VO in our models, we see a discrepancy here. The second row, left column shows a higher metallicity value. The agreement between the two models is very good, as the dominant absorbers for this set of parameters are present in both. The second row, right column shows a higher C/O ratio. Here we start to see another discrepancy between the models, which worsens with increasing C/O ratio. This is due to the differing chemical models. \cite{goyal19} include many more species in their equilibrium model, leading to different abundances for the main absorbers. As a test, we took the abundances from the \cite{goyal19} model and ran the analytical model, which resulted in a very good agreement between the two (not shown). The bottom row, left column shows a higher haze parameter value, which controls the level of the Rayleigh scattering slope. Again, the agreement here is good. Similarly, the bottom row, right column shows a higher cloud value, and the agreement between the two models is good.

\subsection{Sampling Techniques}

Here we describe three possible sampling methods for model grids -- linear sampling, random sampling, and Latin hypercube sampling. We perform a comparison of these methods, with the results shown in Section \ref{sec:sampling_comparison}.

\subsubsection{Linear Sampling}

Traditional grids of models are typically sampled linearly \citep[e.g.][]{allard01,goyal19,marley21}. This involves having one model for each possible combination of parameters, leading to $X^n$ models, where $n$ is the number of parameters and $X$ is the number of values sampled for each one. This is shown in the left panel of Figure  \ref{fig:sampling_schematic}, for an example with 2 parameters, each sampled twice. One of the benefits of a linear grid is that one is able to easily study the effects of a single parameter by comparing consecutive models. Varying only one parameter at a time prevents the effects from multiple parameters becoming entangled. This is extremely useful in forward modelling, where the main goal is to study these effects. 

However, in the field of atmospheric retrieval these grids can prove challenging. Unless strong assumptions are made (e.g. chemical equilibrium), retrievals regularly contain $\sim$10-20 parameters. This quickly escalates the linear grid to a completely unfeasible size. Nevertheless, linear grids are sometimes used with interpolation to perform traditional Bayesian retrievals of exoplanets \citep[e.g.][]{molliere20,miller20,carriongonzalez21}.

\subsubsection{Random Sampling}

In our previous work on machine learning retrievals \citep{marquez-neila18}, we used randomly sampled grids for our training sets. This involves simply drawing each parameter at random from a uniform distribution inside the desired range. Unlike linear sampling, this grid does not allow one to compare models with only one differing parameter. However, for a fixed number of models, the random grid allows for more points in each parameter dimension to be sampled. The right panel of Figure \ref{fig:sampling_schematic} shows an example of a random grid with 4 models. This method proves beneficial for the random forest, which is able to automatically disentangle the effects of each parameter.

\subsubsection{Latin Hypercube Sampling}

\begin{figure*}[t!]
    \centering
    \includegraphics[width=0.9\textwidth]{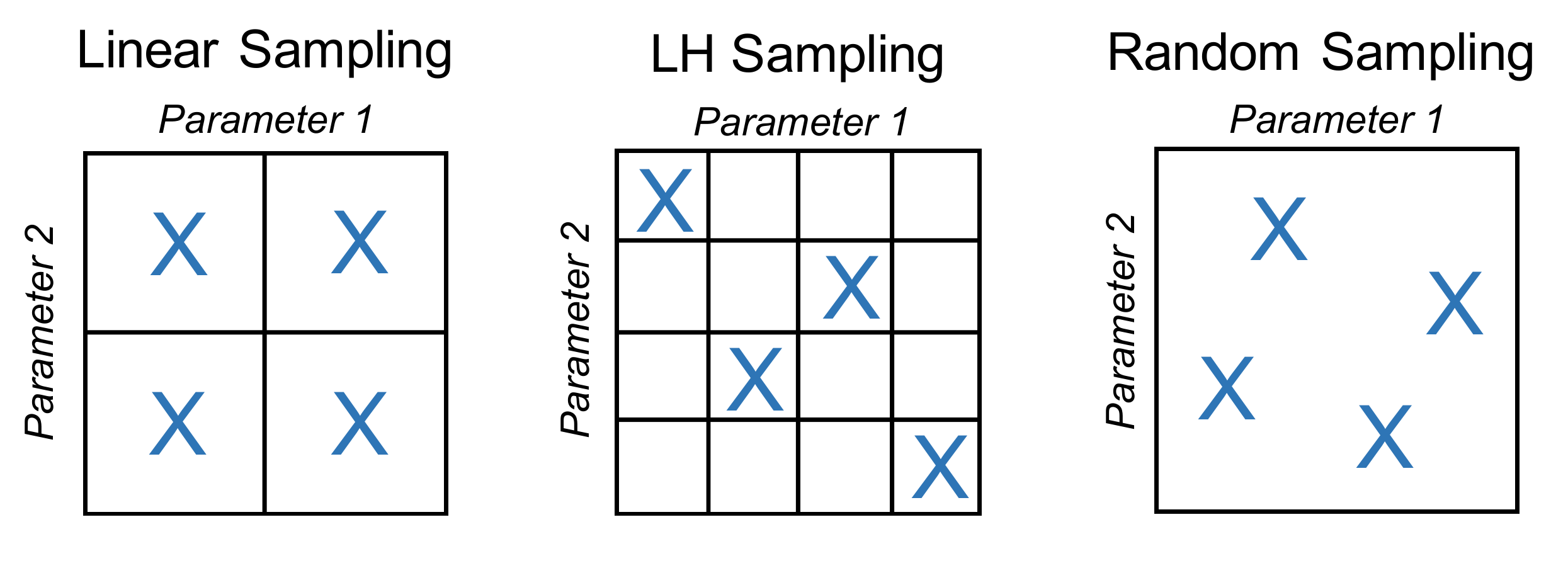}
    \caption{Schematic of different methods for sampling four models from a grid with two parameters. The first panel shows linear sampling, where each parameter has two values and every combination of values is sampled. The second panel shows a Latin square, where the value of each parameter is chosen randomly, but each value is chosen exactly once, allowing for four parameter values each in the grid. The third panel shows completely random sampling, without using a grid of values.}
    \label{fig:sampling_schematic}
\end{figure*}

A common sampling technique in machine learning is Latin hypercube sampling (LHS) \citep{mckay79}. Starting from a square grid with fixed sampling positions, a Latin square has exactly one sample in each row and column. A Latin hypercube is the generalisation of this to higher dimensions. The middle panel of Figure \ref{fig:sampling_schematic} shows an example of a Latin square with 4 models. Latin squares and hypercubes have been used in the design of experiments for almost a century, and provide a method for improving inference from a sparse sampling of high-dimensional space. This has already proven extremely useful in the field of cosmology \citep[e.g.][]{kaufman11,albers19,rogers19,wibking20}, but has yet to be taken advantage of in other fields of astrophysics. 

In comparison to simple random sampling, the key advantage of LHS is that it guarantees a better representation of the real variability of the parameters. Random sampling has no such guarantees. Therefore, for an inference problem with a large number of parameters, LHS typically requires fewer samples than random sampling. See Appendix \ref{sec:lhs} for more details on the history and applications of Latin hypercubes in experiments.

\section{Results}

\subsection{Free Chemistry Grid}

\subsubsection{Sampling Comparison}
\label{sec:freechem_samplingcomp}

To compare the different methods, we created training sets using each of the three sampling techniques, and then trained and tested a corresponding random forest. For a fair comparison, we assumed a fixed number of models can be generated, such that all three training sets have the same size. This is also a realistic situation in which one is limited by computation time. We considered five different models, with two, four, six, eight, and eleven parameters. For linear sampling, we sampled each parameter an equal number of times, given by $X$, and calculated as the highest integer such that $X^n\sim$100,000. This results in the training set sizes shown in Table \ref{tab:training_size}. These sizes are kept the same for the random sampling and LHS training sets. The parameters in the linear and LHS grids are evenly spaced inside the prior range (shown in Table \ref{tab:priors}), while the parameters in the random grids are drawn randomly from a uniform distribution across this range.

\begin{table}[]
    \centering
    \begin{tabular}{|c|c|c|}
        \hline
        \# Parameters & \# Parameter Samples & Training Set Size \\
        \hline
        2 & 316 & 99856 \\
        4 & 17 & 83521 \\
        6 & 6 & 46656 \\
        8 & 4 & 65536 \\
        11 & 3 & 177147 \\
        \hline
    \end{tabular}
    \caption{Table showing the training set size for our 2-, 4-, 6-, 8-, and 11-parameter models.}
    \label{tab:training_size}
\end{table}

\begin{figure*}
    \centering
    \includegraphics[width=0.9\textwidth]{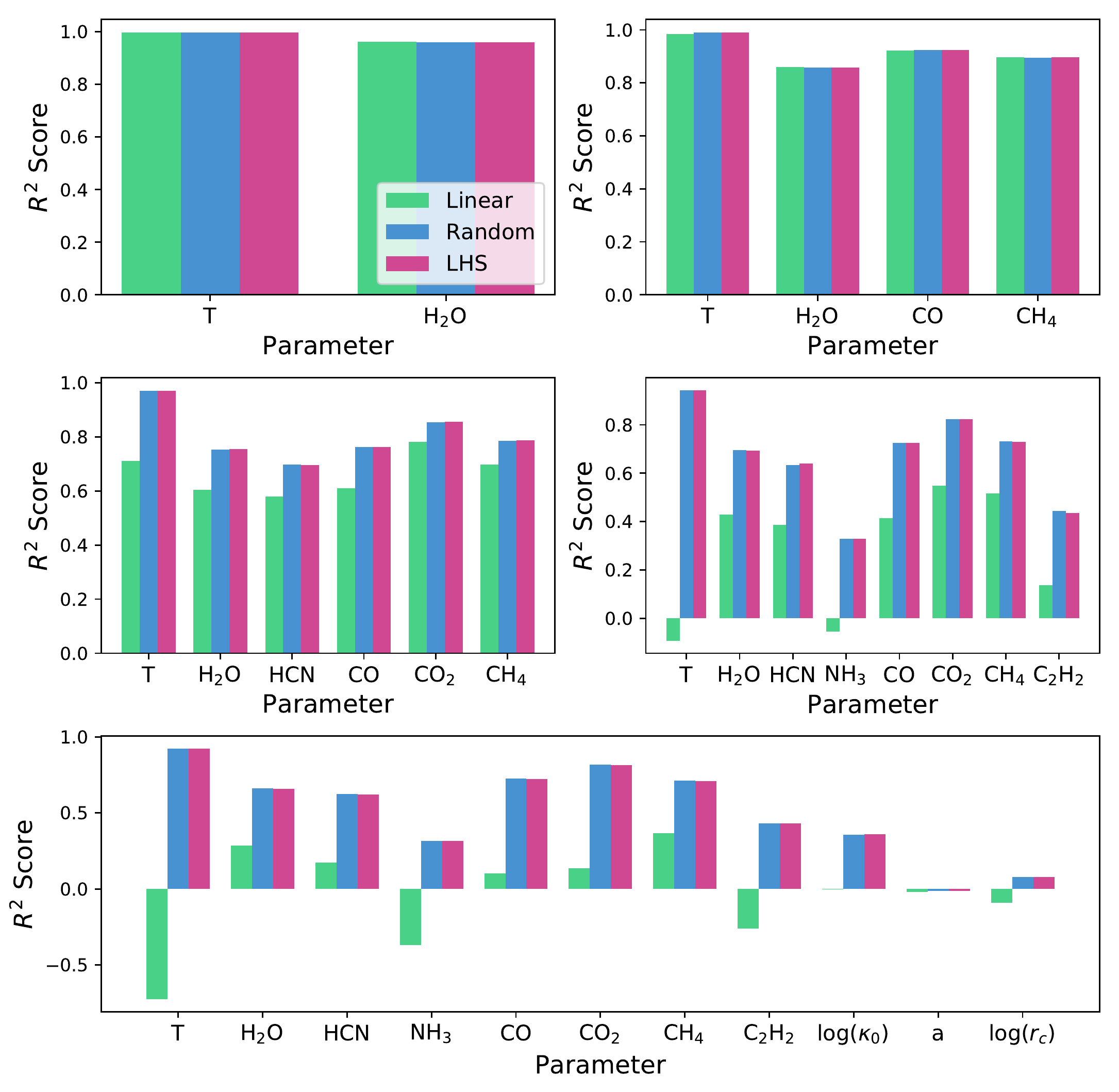}
    \caption{$R^2$ scores for each parameter in each of the free chemistry models, using the three different grid-sampling methods -- linear sampling, random sampling and Latin hypercube sampling.}
    \label{fig:r2_barchart_freechem}
\end{figure*}

For each model, a forest is trained on each of the differently sampled grids. Due to its ability to make fast predictions, the forest can be tested on a large set of models, spanning the whole parameter space. We randomly generated a testing set of 10,000 models, keeping it the same across each sampling case. Once tested, the forest provides a useful predictability analysis, by calculating the coefficient of determination ($R^2$ score) between the real and predicted values of each parameter. The $R^2$ score varies from $-1$ to 1, where values near unity indicate strong anticorrelations and correlations, respectively, between the real and predicted values of a given parameter. Figure \ref{fig:r2_barchart_freechem} shows the coefficient of determination for each parameter and sampling technique in the different models. 

For the 2- and 4-parameter models, all three training sets perform comparably well. This is expected as even in the linear case, each parameter is still sampled sufficiently densely for the forest to learn its effects. For the 6-parameter model, the linear case starts to drop in performance, when compared with the random and LHS cases. For the 8-parameter model, the linear case shows an extremely poor predictability for certain parameters, including temperature, which had previously been relatively easy to predict. This is because each parameter is only sampled 4 times, providing the forest with little information on the parameters' effects on the spectrum. The random and LHS cases show fairly low $R^2$ scores for the newly added parameters (NH$_3$ and C$_2$H$_2$), but the other parameters remain fairly well-predicted. The lower $R^2$ scores for NH$_3$ and C$_2$H$_2$ is likely due to their lack of strong, distinctive molecular features, and degeneracies with the other molecules. The 11-parameter model shows an even more extensive difference, with several parameters in the linear grid dropping into negative predictability. 

Generally these results are unsurprising, as it is trivial that parameters sampled only 3 or 4 times in the training set will be hard to retrieve. However, these linear grids are often used with an interpolation scheme in traditional Bayesian retrievals \citep[e.g.][]{molliere17,miller20,carriongonzalez21}, and these results highlight issues that can arise from this. One interesting takeaway from these results is that the $R^2$ score for each parameter in the random and LHS models remains at a very similar value across each model, as more parameters are added. This suggests that the addition of extra parameters does not necessarily require a higher number of samples in the training set. However, it's possible that this could be due to the distinct effects each parameter has on the spectrum in this specific case, as more interconnected parameters could be harder to disentangle. So far, we see very little difference between the random and LHS cases, most likely because the number of parameters is still relatively low. 

\subsubsection{Mock Retrieval}

\begin{figure*}
    \centering
    \includegraphics[width=0.9\textwidth]{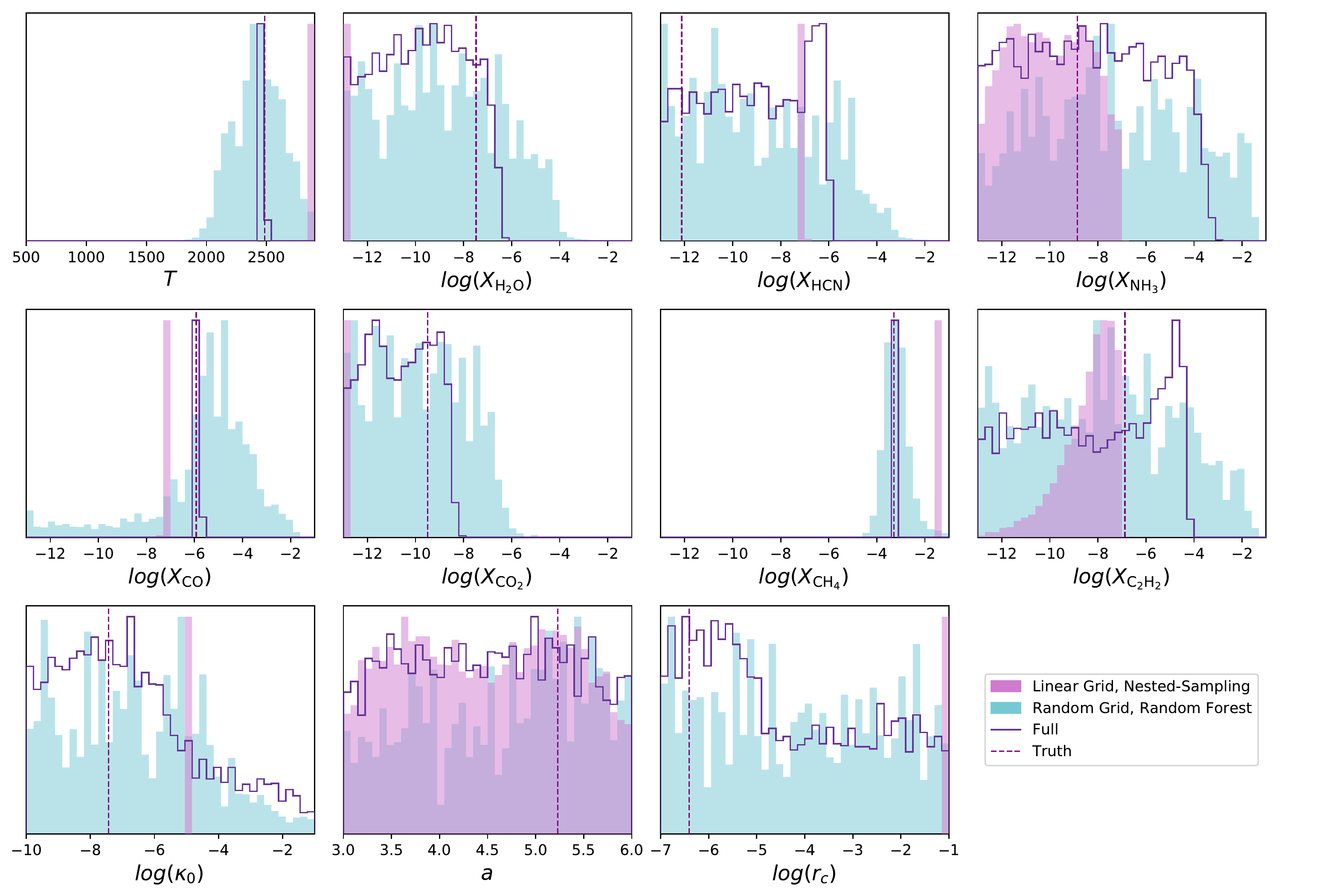}
    \caption{Retrieval of a mock spectrum from the 11-parameter model using different methods. This shows the retrievals using nested-sampling, interpolating on the linear grid, and the random forest trained on the random grid, compared with the full retrieval using nested-sampling with the analytical model computed on-the-fly. The dashed purple lines show the true parameter values for the spectrum.}
    \label{fig:mock_retrieval_freechem}
\end{figure*}

One potential advantage of a linearly spaced grid is that it allows for easy interpolation, and can therefore be used in a traditional, Bayesian retrieval with an MCMC or nested-sampling. This is beneficial when the forward model is quite slow, as a traditional retrieval needs to compute models on-the-fly, typically tens of thousands of times for a single run. By computing a grid in advance, the computational burden is shifted offline, and the models can be re-used for multiple retrievals. To study this, we ran several retrievals on a simulated spectrum using our free chemistry models. Figure \ref{fig:mock_retrieval_freechem} shows three retrievals performed on the same simulated spectrum for the 11-parameter model. The first uses nested-sampling by interpolating on the linear grid. The second uses the random forest trained on the random grid. The third uses nested-sampling with the full analytical model. 

The key difference between the full analytical retrieval and the random forest trained on the random grid is the width of the posteriors. For the tightly constrained parameters in the full retrieval, such as temperature, CO, and CH$_4$, the forest's posterior generally peaks in the same place, but with a wider distribution. This is due to nested-sampling's ability to hone in on a small part of parameter space. For the parameters with an upper bounded full retrieval posterior, such as H$_2$O, HCN, NH$_3$, CO$_2$, and C$_2$H$_2$, the upper limit for the forest's posterior is about 2 dex higher. For the completely unconstrained cloud parameters, the forest's posteriors are comparable. Improvements can be made on the forest's posteriors by adjusting parameters in the forest such as the number of trees or tree depth, increasing the size of the training set, or incorporating a likelihood into the posterior computation \citep{nixon20}. However, the latter negates one advantage of the traditional likelihood-free random forest, which does not rely on assuming a functional form of the likelihood.

Unsurprisingly, the nested-sampling retrieval using interpolation on the linear grid performs poorly due to the sparse sampling of the parameters. It would not be appropriate to use interpolation for a grid with this many dimensions.

\subsection{Goyal Grid}

\subsubsection{Sampling Comparison}
\label{sec:sampling_comparison}

Using the analytical model described in Section \ref{sec:analytical_model}, we created four different versions of the Goyal grid. The first two are linearly spaced grids of different sizes. We started by following the parameter spacing from \cite{goyal19}, described in Section \ref{sec:goyal_models}. This consists of 24 temperatures, 4 gravities, 6 metallicities, 4 C/O ratios, 4 haze parameters, and 4 cloud parameters, leading to a total of 36,864 models. Next we created a sparser version of this linear grid, sampling fewer parameter values. This consists of 5 temperatures (300, 900, 1500, 2100, 2600 K), 3 gravities (5, 14, 50 ms$^{-2}$), 4 metallicities (0.1, 1, 50, 200), 3 C/O ratios (0.35, 0.63, 1.0), 3 haze parameters (1, 31.6, 1100), and 3 cloud parameters (0, 0.13, 1.0), leading to a total of 1620 models. A summary of the parameter spacing for both versions of the linear grid is show in Table \ref{tab:goyal_parameters}. Note that in the analysis we converted the gravity, metallicity, and haze parameters to log quantities. 

\begin{table*}
\begin{center}
\caption{Parameter values for the grid from \cite{goyal19}, and the values in our sparsely sampled linear grid.}
\label{tab:goyal_parameters}
\vspace{0.1in}
\begin{tabular}{|l|c|c|}
\hline
    Parameter & Goyal grid & Sparse grid \\
    \hline
    T (K) & (300-2600), in steps of 100 & (300, 900, 1500, 2100, 2600) \\
    g (ms$^{-2}$) & (5, 10, 20, 50) & (5, 14, 50) \\
    metallicity (x solar) & (0.1, 1, 10, 50, 100, 200) & (0.1, 1, 50, 200) \\
    C/O & (0.35, 0.56, 0.7, 1.0) & (0.35, 0.63, 1.0) \\
    Haze & (1, 10, 100, 1100) & (1, 31.6, 1100) \\
    Cloud & (0, 0.06, 0.2, 1.0) & (0, 0.13, 1.0) \\
    \hline
\end{tabular}
\end{center}
\end{table*}

The second two grids use different sampling methods. First is a randomly sampled set of 1620 models, created by drawing each parameter from a uniform distribution (or log-uniform for metallicity, gravity, and the haze parameter) in the same range as in \cite{goyal19}. The final grid is a set of 1620 models using LHS, where each parameter dimension has 1620 evenly-spaced points (or even in log-space for metallicity, gravity, and the haze parameter) in the same range as in \cite{goyal19}.

\begin{figure}
    \centering
    \includegraphics[width=\columnwidth]{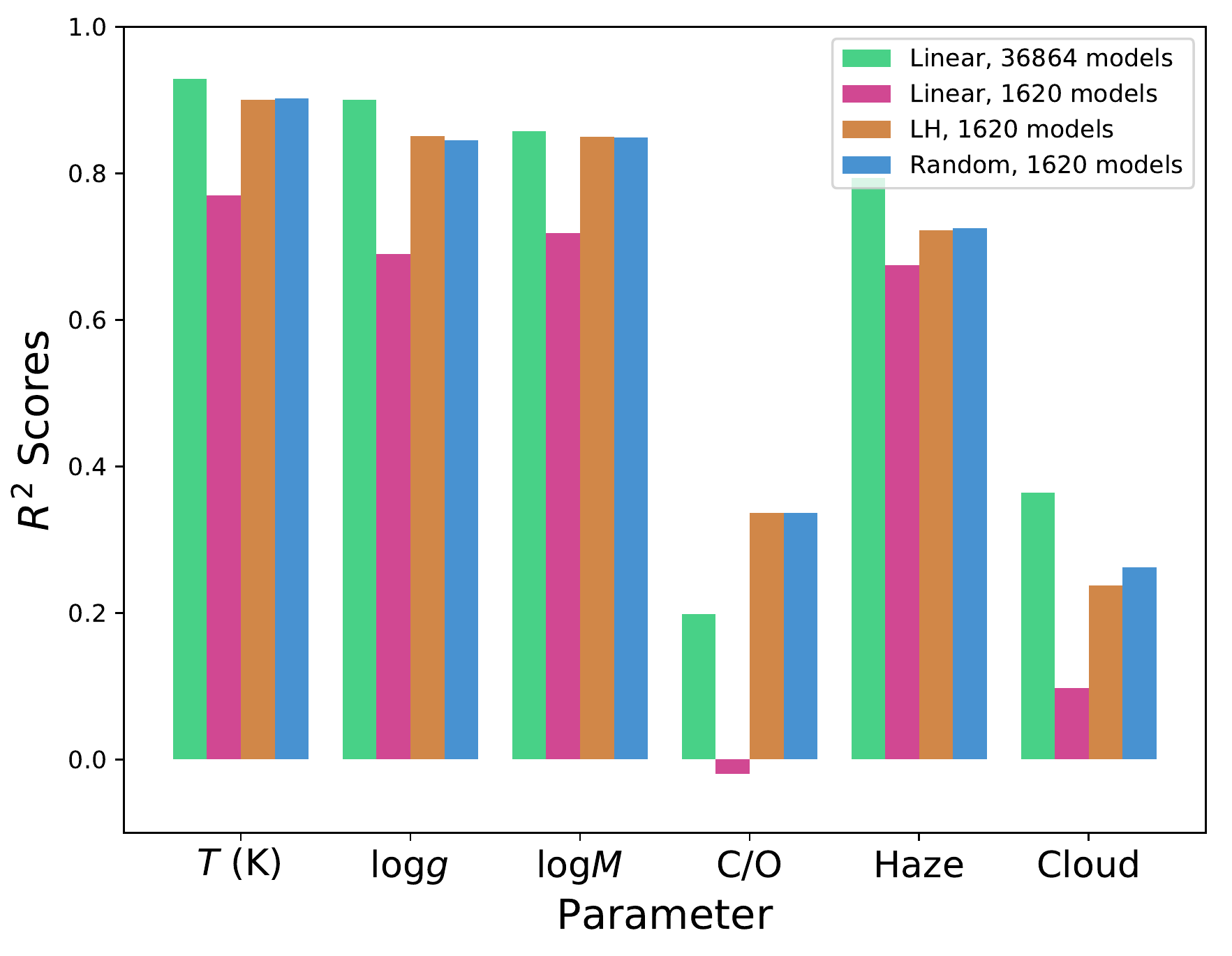}
    \caption{$R^2$ scores for each parameter in the Goyal model, using the three different grid-sampling methods -- linear sampling, random sampling and Latin hypercube sampling. Linear sampling is tested for a dense grid of 36,864 models, and a sparse grid of 1620 models.}
    \label{fig:r2_barchart_goyal}
\end{figure}

We trained a random forest on each of the four grids, and then tested them on a randomly generated set of 5000 models. Figure \ref{fig:r2_barchart_goyal} shows the $R^2$ scores for each parameter across the different grids. 

For every parameter, the sparse linear grid is out-performed somewhat by all other grids, including the random and LHS grids that contain the same number of models. For most parameters, the denser linear grid, the random grid, and the LHS grid give comparable results, despite differing in size by more than a factor of 20. Although the predictability of the C/O ratio is significantly higher for the random and LHS grids, this is actually due to the uneven spacing adopted for the linear grid. In \cite{goyal19}, the C/O ratio takes the values 0.35, 0.56, 0.7 and 1.0, leaving a larger gap at the higher values. Due to the non-linear effect of the C/O ratio on the spectrum, it proves challenging for the forest to accurately interpret models in this high C/O range. In contrast, for temperature even the sparse linear grid performs relatively well. In fact, all four well-predicted parameters (i.e. temperature, gravity, metallicity, and haze) have comparable $R^2$ scores across all models, to within $\sim0.1$. This contrasts the results from Section \ref{sec:freechem_samplingcomp}, but could be explained by the relatively low number of parameters in the model and the linear effects of these four parameters on the spectrum. This motivates a differently structured grid, with denser sampling for parameters with highly non-linear effects on the spectrum. 

Since LHS guarantees a better representation of the real variability of the parameters, we might expect it to outperform random sampling. The fact that the results in the bar chart in Figure \ref{fig:r2_barchart_goyal} look comparable between the two could be due to the relatively low number of parameters. Perhaps a higher number of dimensions could lead to a divergence in their performances.

\subsubsection{Mock Retrieval}

\begin{figure*}
    \centering
    \includegraphics[width=0.9\textwidth]{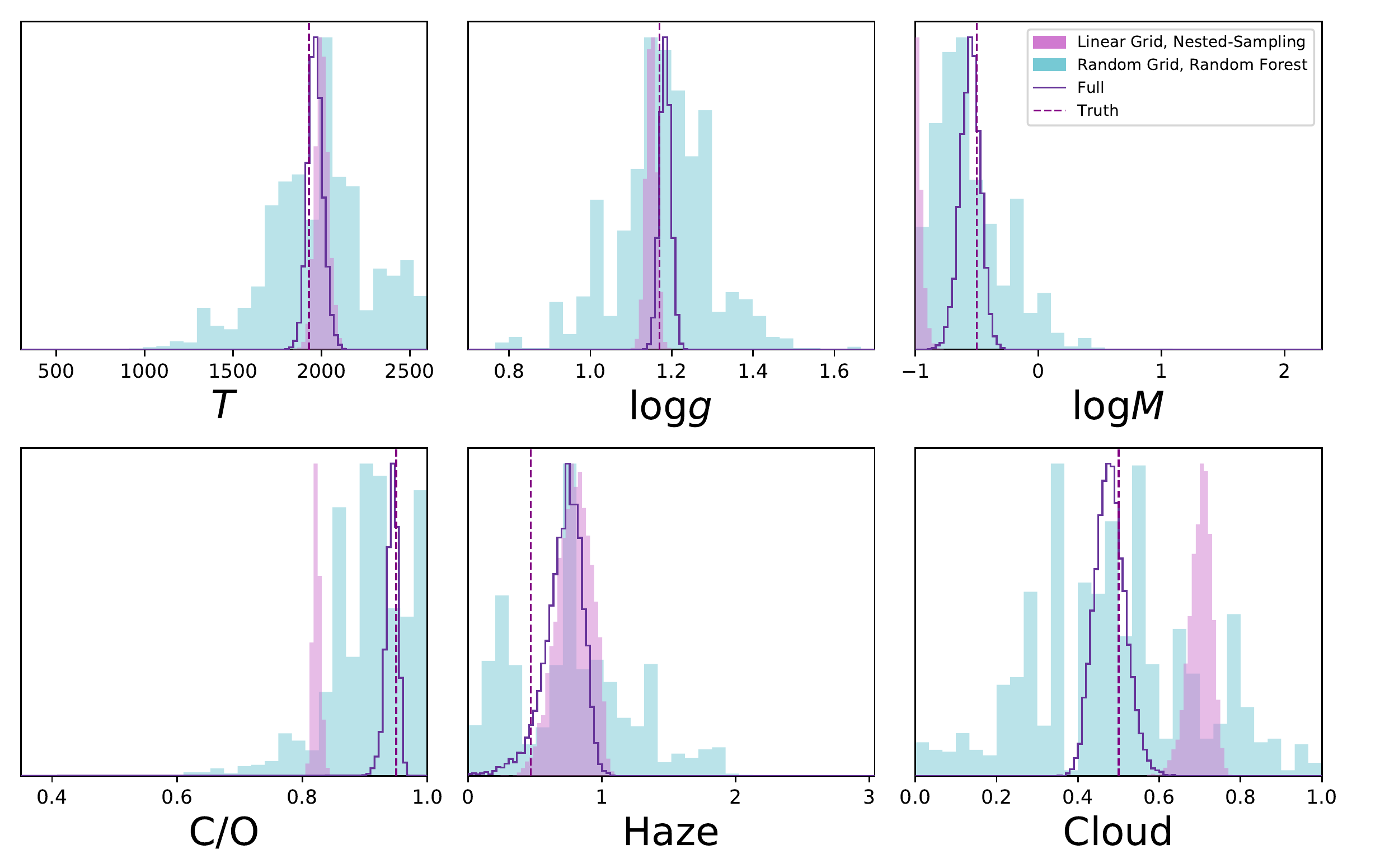}
    \caption{Retrieval of a mock spectrum using different methods. This shows the retrievals using nested-sampling, interpolating on the dense linear grid, and the random forest trained on the dense random grid, compared with the full retrieval using nested-sampling with the analytical model computed on-the-fly. The dashed purple lines show the true parameter values for the spectrum.}
    \label{fig:mock_retrieval_goyal}
\end{figure*}

We also ran mock retrievals using the Goyal grids. One example is shown in Figure \ref{fig:mock_retrieval_goyal}. This figure shows three retrievals. The first uses nested-sampling with interpolation on the dense linear grid. The second uses the random forest, trained on a randomly sampled grid of 36,864 models (i.e. the same size as the dense linear grid). The third uses nested-sampling with the full analytical model, computed on-the-fly. In contrast to Figure \ref{fig:mock_retrieval_freechem}, the linear interpolation retrieval does not perform so badly. Here, the temperature, gravity, and haze posteriors are very similar to the full retrieval, with only minor offsets. Of course this is due to denser sampling of the parameters, although the gravity and haze, for example, only sample one extra point than the parameters in the free chemistry model from Figure \ref{fig:mock_retrieval_freechem}. It could be that the effects of these parameters are more linear than those in the free chemistry model. The poor performance of the linear interpolation retrieval is seen again in the metallicity, C/O ratio, and the cloud parameters. This is likely due to the more non-linear effects of these parameters, and the uneven spacing (for the C/O ratio in particular).  

The random forest posteriors exhibit a similar behaviour as in Figure \ref{fig:mock_retrieval_freechem}, with wider, less constrained distributions, but with the same peak location. This effect is exaggerated for the cloud parameter, for which the forest's posterior encompasses the entire range of values. This is consistent with the R$^2$ scores in Figure \ref{fig:r2_barchart_goyal}, which shows that the forest struggles to accurately retrieve the cloud parameter.

\subsubsection{Information Content of the Goyal Grid}

\begin{figure*}
    \centering
    \includegraphics[width=\textwidth]{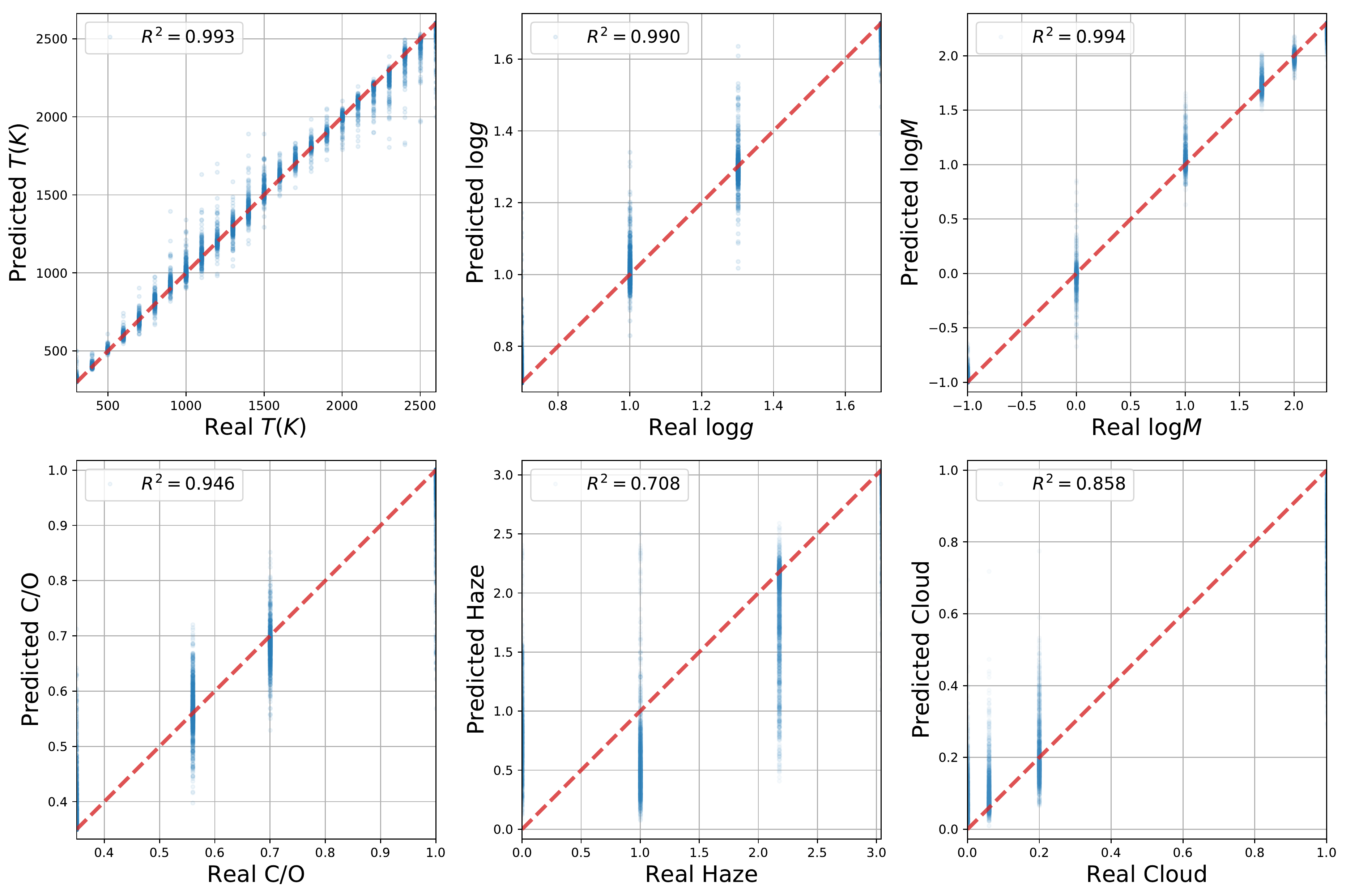}
    \caption{Real vs predicted values for the random forest trained on the models from \cite{goyal19}, binned down to match the resolution and wavelength coverage of JWST NIRSpec Prism. The $R^2$ score varies from $-1$ to 1, where values near unity indicate strong anticorrelations and correlations, respectively, between the real and predicted values of a given parameter.}
    \label{fig:goyal_predvreal}
\end{figure*}

We also trained a forest on the original grid of models from \cite{goyal19}. To be consistent, we binned these models down to the same resolution and wavelength range, and added the same noise level, as in Section \ref{sec:analytical_model}. We then randomly selected 6864 models to be the testing set, leaving 30,000 for training. Figure \ref{fig:goyal_predvreal} shows the predicted vs real values for this forest. Although the $R^2$ scores are high for all 6 parameters, an important caveat is that the testing set contains the same evenly spaced parameter values as the training set. This makes it significantly easier for the forest to predict the correct answer. The sparse testing set provides no information about how the forest performs on spectra with parameters in between these values. We would need to generate our own \texttt{ATMO} models with randomly chosen parameters in order to test this forest's ability to generalise.

\begin{figure*}
    \centering
    \includegraphics[width=\textwidth]{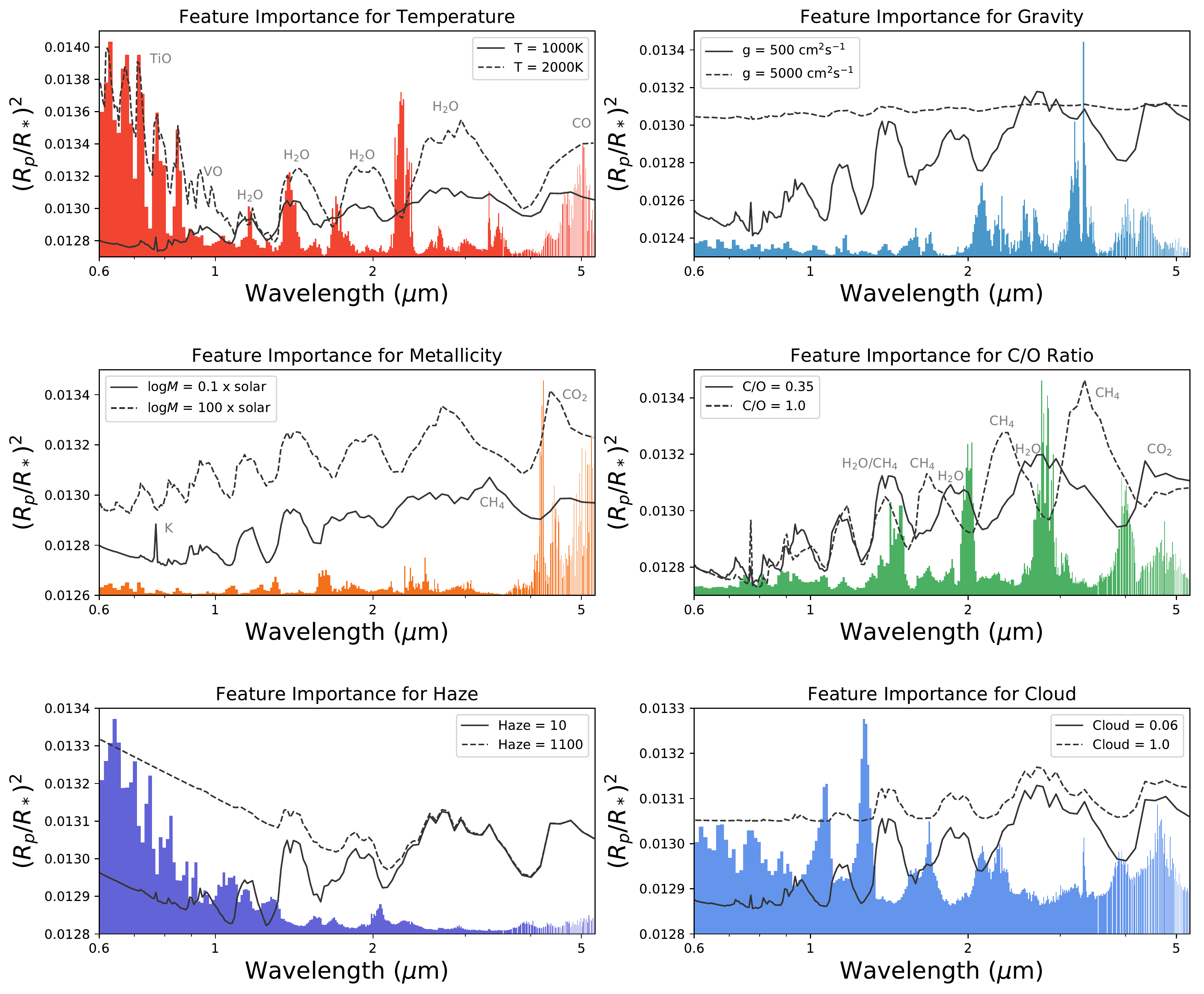}
    \caption{Relative feature importance of each wavelength point for each parameter of the model. Plotted over the top are two models with high and low values of the corresponding parameter, showing examples of the behaviour of different spectra. This allows for the comparison between spectral features and their relative importance for each parameter. The y-axes correspond to the transit depth for the models. The feature importance values add up to unity, though their values are not shown.}
    \label{fig:feature_importance}
\end{figure*}

In addition, we computed the ``feature importance" for this forest. This determines the information content of each spectral point with respect to each parameter in the retrieval. Figure \ref{fig:feature_importance} shows the results for the original Goyal grid. In this figure, the feature importance is shown against two spectra with a high and low value of the relevant parameter, to provide context. Some aspects of the feature importance are intuitive, such as the haze parameter drawing most information from the bluer wavelengths, where the Rayleigh slope is visible. Parameters that are less well-constrained typically have a slightly more uniform feature importance across all spectral points, as is seen for the cloud parameter. Temperature has major peaks at the TiO features, implying that this species performs well as a thermometer for the atmosphere. The C/O feature importance appears to peak around the water features at 2 and 3 $\mu$m, which are present in spectra with lower C/O ratios. These features coincide with troughs in the methane opacity, which dominates the high-carbon spectra, making these areas good indicators of the C/O ratio. The same behaviour in the feature importance is found for the forests trained on the analytical model, using both the linear and random grids (not shown).

\section{Conclusion}

We generated differently sampled grids for two types of atmospheric models, and compared their performance using different types of analysis. For our free chemistry model, we found that random and Latin hypercube sampling outperformed linear sampling for our 8- and 11-parameter models, but obtained comparable results for our 2-, 4-, and 6-parameter models (Figure \ref{fig:r2_barchart_freechem}). Our free chemistry mock retrieval for the 11-parameter model clearly demonstrated that linear interpolation is not appropriate for high dimensional models, as expected (Figure \ref{fig:mock_retrieval_freechem}). For our models simulating those of \cite{goyal19}, we found that the difference between the linear and random or LH sampling was less significant, particularly for parameters with generally high predictability, likely due to the lower dimensionality of the model. Our mock retrieval showed that the linear interpolation retrieval performed well for several parameters, but struggled with parameters with non-linear effects, such as the C/O ratio. 

These results warn against the use of linear interpolation of pre-computed linear model grids for atmospheric retrievals. They also demonstrate the known results that random and Latin hypercube sampling enable inference from a sparsely sampled parameter space. We did not find an improvement in LH sampling over random sampling, likely due to the relatively low dimensionality of our models. Although they aren't ideal for building into a retrieval, linearly sampled grids have their own advantages. For example, they enable one to easily compare spectra differing in only one parameter, allowing for the clear analysis of individual parameter effects \citep[e.g.][]{goyal19}. Furthermore, the feature importance for the linear grid from \cite{goyal19} provided extremely useful analysis of the information content of the spectra, although this could also be performed on the other grids.

Therefore, the answer for which type of sampling to use when constructing a model grid depends heavily on the intended use of the grid. For retrievals, in particular for machine learning, randomly sampled grids are likely to provide better results, with higher predictability for each model parameter, and require far fewer models. However, for analysis of the model physics and spectral sensitivity, a linearly sampled grid is preferable, due to the ease of model comparisons. Information content analysis can be performed in either case, providing similar results, beneficial for observing proposals or even informing future telescope development.

\begin{acknowledgements}
We thank Daniel Kitzmann for helpful discussions about the model, and Pablo M\'{a}rquez-Neila for advice about the random forest. We acknowledge financial support from the Swiss National Science Foundation (via a Postdoc Mobility grant to CF; grant number P500PT\texttt{\char`_}203110), the European Research Council (via a Consolidator Grant to KH; grant number 771620), the PlanetS National Center of Competence in Research (NCCR), the Center for Space and Habitability (CSH) and the Swiss-based MERAC Foundation. Finally, we thank the anonymous referee for helping to improve this work.
\end{acknowledgements}
\software{
        \texttt{HELIOS-K} \citep{grimm15}, 
        \texttt{scikit.learn} \citep{pedregosa11}, 
        \texttt{PyTorch} \citep{paszke19}, 
        \texttt{Adam} \citep{kingma14}, 
        \texttt{PyMultinest} \citep{buchner14}, 
        \texttt{Astropy} \citep{astropy18}, 
        \texttt{numpy} \citep{harris20}, 
        \texttt{scipy} \citep{virtanen20}, 
        \texttt{matplotlib} \citep{hunter07}.
        }

\bibliographystyle{aasjournal}
\bibliography{mybib}

\appendix
\section{Latin Hypercubes in Computer Experiments}
\label{sec:lhs}

The use of Latin squares in the design of experiments, particularly in agriculture, dates back almost a century \citep[][]{fisher35}. Its modern applications to computer experiments enables inference from a sparse coverage of high-dimensional parameter space \citep[][]{santner03,fang06,kleijnen15}. However, the use of Latin squares, and the higher dimensional Latin hypercubes, has been relatively limited in astrophysics, with the majority of applications in cosmology. \cite{kaufman11} use Latin hypercubes to improve the efficiency of emulators, and apply it to photometric redshifts of galaxies. More recently, \cite{wibking20} use Latin hypercubes for their emulation framework modelling galaxy clustering, while \cite{albers19} implement them in the training of a neural network to speed-up Einstein-Boltzmann Solvers in cosmological simulations. \cite{rogers19} also use Latin hypercubes in their emulator for the Lyman-alpha forest, for example.

There has been a great amount of work in the statistics community on the optimisation of Latin hypercube designs (LHDs) to improve their efficiency and apply them to ensemble models, such as orthogonal LHDs \citep[e.g.][]{sun10}, sliced LHDs \citep[e.g.][]{ba15,qian12}, and progressive LHS \citep[e.g.][]{sheikholeslami17}. There are also many packages in the \texttt{R} public domain software environment for using LHS for computer experiments (e.g. \texttt{lhs} \citep{carnell20}; \texttt{DiceDesign} and \texttt{DiceEval} \citep{dupuy15}; \texttt{DiceKriging} \citep{roustant12}; \texttt{DiceView} \citep{ritchet20}; \texttt{tgp: Bayesian Treed Gaussian Process Models} \citep{gramacy07,gramacy10}).

\label{lastpage}

\end{document}